\documentstyle[12pt,fleqn]{article}
\def\R{I\kern-.2emR}
\def\1{\bar{1}}
\def\One{\mbox{{\sf 1}\kern-0.25em{\bf l}}}
\def\beq{\begin{equation}}
\def\eeq{\end{equation}}
\def\str{\rule{0pt}{2.4ex}}

\begin{document}
\baselineskip 18pt plus 2pt
\vspace*{15mm}
\begin{center}
{\Large Kochen-Specker theorem for 8-dimensional space} \\[2cm]
Michael Kernaghan\\[7mm]
{\sl Department of Philosophy, University of Western Ontario, London
N6A 3K7, Canada}\\[1cm]
and\\[1cm]
Asher Peres\\[7mm]
{\sl Department of Physics, Technion -- Israel Institute of
Technology, 32 000 Haifa, Israel}\\[1cm]
\end{center}\vfill

\noindent{\bf Abstract}\bigskip

A Kochen-Specker contradiction is produced with 36 vectors in a real
8-dimensional Hilbert space. These vectors can be combined into 30
distinct projection operators (14 of rank 2, and 16 of rank 1). A
state-specific variant of this contradiction requires only 13 vectors,
a remarkably low number for 8 dimensions.
\vfill\newpage

The Kochen-Specker theorem [1] asserts that, in a Hilbert space with a
finite number of dimensions, $d\geq3$, it is possible to produce a set
of $n$ projection operators, representing yes-no questions about a
quantum system, such that none of the $2^n$ possible answers is
compatible with the sum rules of quantum mechanics. Namely, if a subset
of mutually orthogonal projection operators sums up to the unit matrix,
one and only one of the answers is yes. The physical meaning of this
theorem is that there is no way of introducing noncontextual ``hidden''
variables [2] which would ascribe definite outcomes to these $n$ yes-no
tests. This conclusion holds irrespective of the quantum state of the
system being tested.

It is also possible to formulate a ``state-specific'' version of this
theorem, valid for systems which have been prepared in a known pure
state. In that case, the projection operators are chosen in a way
adapted to the known state. A smaller number of questions is then
sufficient to obtain incompatibility with the quantum mechanical sum
rules.  An even smaller number is needed if strict sum rules are
replaced by weaker probabilistic arguments [3,4].

The original proof by Kochen and Specker [1] involved projection
operators over 117 vectors in a 3-dimensional real Hilbert space \R$^3$.
A simple proof with 33 vectors was later given by Peres [5], who also
reported an unpublished construction by Conway and Kochen, using only 31
vectors [6]. A proof with 20 vectors in \R$^4$ was recently given by
Kernaghan [7]. Here, we consider the Kochen-Specker theorem in \R$^8$. A
state-independent proof is produced with 36 vectors, which can be
collected into 30 distinct projection operators. A state-specific proof
involves only 13 vectors, and achieves the lowest value of the ratio
$n/d$ that has been obtained so far.

Our construction is based on Mermin's remark [8] that, for any three
spin-$1\over2$ particles, the four operators

\beq A=\sigma_{1z}\otimes\sigma_{2z}\otimes\sigma_{3z},\eeq
\beq B=\sigma_{1z}\otimes\sigma_{2x}\otimes\sigma_{3x},\eeq
\beq C=\sigma_{1x}\otimes\sigma_{2z}\otimes\sigma_{3x},\eeq
\beq D=\sigma_{1x}\otimes\sigma_{2x}\otimes\sigma_{3z},\eeq
commute. Moreover, their product is proportional to the unit matrix,

\beq ABCD=-\One. \eeq
Mermin's three particle system is the simplest one that illustrates {\em
both\/} Bell's theorems, on nonlocality [9] and contextuality [10].

Each one of the five equations above involves a complete set of
commuting operators, and therefore determines a complete orthogonal
basis in \R$^8$ (complex numbers are not needed here, because $\sigma_x$
and $\sigma_z$ are real matrices). The five orthogonal octads generated
by the above equations are listed in the first column of Table 1. To
simplify typography, each vector was given norm 2 (this avoids the use
of fractions) and the symbol \=1 means $-1$. The components of each
vector are written as a horizontal array, rather than the usual ``column
vector.'' The basis used is the direct product of the bases where each
$\sigma_z$ is diagonal. For example, the vector with components 00002000
is the eigenvector for which $\sigma_{1z}$ has eigenvalue $-1$, and
$\sigma_{2z}$ and $\sigma_{3z}$ have eigenvalue 1. This is most easily
seen by using binary digits, 0 and 1, for labelling the ``up'' and
``down'' components of a spinor, respectively, and combining them into
binary numbers, 000, 001, \ldots, 111, for labelling components of
vectors in \R$^8$. Thus, the vector 00002000 has the physical meaning
stated above because its only nonvanishing component is the 100th one
(that is, the fifth one, in binary notation).

It is easy to see that the 40 projection operators on these 40 vectors
provide an example of the Kochen-Specker contradiction. Indeed,
associating values 0 (no) and 1 (yes) to the vectors in a basis (with a
single 1, of course) amounts to selecting one of these eigenvectors, and
the latter indicates which ones of the four commuting operators which
generate that basis have value 1, and which ones have value $-1$. Such a
mapping of each operator on one of its eigenvalues cannot be done for
all of them in Eqs.~(1--5): it would lead to an inconsistency, because
of the minus sign in (5). This argument is readily generalized to a
larger number of spin-$1\over2$ particles, and provides a proof of the
Kochen-Specker theorem in a real Hilbert space with $2^n$ dimensions.
However, fewer than $2^n$ vectors are actually needed for the proof, as
we shall see.

First, we note that the above set of 40 vectors has a high degree of
symmetry. (Unfortunately, we have not been able to determine its
invariance group. We asked several well known experts, who also did
not find it.) A direct inspection, best done by computer, shows that
each vector is orthogonal to 23 other ones (7 in the same basis, and 4
in each one of the four other bases), and it makes a 60$^\circ$ or
120$^\circ$ angle with each one of the 16 remaining vectors.  It is
possible to construct with the 40 vectors 25 distinct orthogonal octads
(each vector appears in 5 octads).  Eleven of these octads are listed in
the remaining columns of Table 1 (the 14 other octads are not needed for
the proof and have not been listed).

It is seen that there are four vectors, namely those with components
20000000, 00001111, 001\=1001\=1, and 10\=10\=1010, that do not appear
in the 11 octads (they appear of course in the 14 unlisted octads).
These four vectors are mutually orthogonal, and they belong to four
different of the original octads. It is easily seen that there are 1280
different ways of choosing four vectors with these properties (that is,
of choosing which are the 11 octads, out of 25, that appear in Table
1).

It is also seen that each one of the 36 remaining vectors appears in
Table 1 either twice or four times. Let us now consider the projection
operators over these 36 vectors.  According to quantum mechanics, each
projection operator corresponds to a yes-no question: the eigenvalues 1
and 0 mean yes and no, respectively. Each orthogonal octad defines 8
commuting projection operators (that is, 8 compatible questions) which
sum up to the unit matrix. This means that if an experimental test is
actually performed for these 8 questions, the answer is yes to one, and
only one of them. The value 1 is thereby associated with one of the
eight vectors of each octad, and the value 0 with the others.

If nothing is known of the state of the system, quantum mechanics is
unable to predict which vector will get the value 1. A natural question
is whether there could be a more complete theory, such that the value
associated with each vector would be determined by ``hidden variables.''
Table 1 readily shows that this goal cannot be achieved. Indeed, the sum
of values in each octad is always 1, therefore the sum of values for the
11 octads in the table is 11, which is an {\em odd\/} number. On the
other hand, each vector (tentatively associated with a value which is
either 0 or 1) appears either twice or four times in the table (with the
{\em same\/} value), thus contributing 2 or 4 (an {\em even\/} number)
to the sum of values.  We have reached a contradiction. This is the
proof of the Kochen-Specker theorem in \R$^8$.

Furthermore, our 36 incompatible propositions can be combined into a
smaller number, namely 30 distinct ones, by using projection operators
of rank~2 on the {\em planes\/} spanned by some pairs of vectors. Table
2 shows how 20 of the vectors can be combined into 14 planes. The
vectors on each line are mutually orthogonal. Each plane is spanned by
two adjacent vectors on the same line (for example, 02000000 and
00000002). When such a pair of vectors occurs in any of the 11 octads
listed in Table~1 (that is, in one of the columns) this vector pair
should be replaced by the corresponding plane. Once this is done, each
column lists planes and unpaired vectors, all of which are mutually
orthogonal.  They correspond to commuting projection operators of rank 1
or~2, which sum up to~\One.  Therefore, if ``experimental'' (that is,
counterfactual) values 0 or~1 are attributed to them, these values sum
up to 11, exactly as before. On the other hand, each one of the
remaining 16 vectors (those not used in Table~2) appears twice in the
new version of Table~1, and each of the 14 planes appears 2 or~4 times
-- always an even number. We thus reach the same Kochen-Specker
contradiction as before.

Until now, we assumed nothing about the state of the quantum system. If
that state is known with certainty, it becomes possible to find a much
smaller number of propositions which lead to a Kochen-Specker
contradiction. For instance, let the quantum system be prepared in the
pure state 100\=10\=1\=10 (this does not restrict the generality of this
discussion in any way, because it is always possible to choose the basis
in \R$^8$ in such a way that any given pure state be represented by a
vector with these components). We can now discard from Table~1 that
vector, whose associated value is~1, by definition, and all the vectors
ortho\-gonal to it, whose associated values are zero. Only 7 non-empty
columns, with 13 different vectors, remain in Table~1. In each column,
there are four mutually orthogonal vectors, which span a subspace
containing the known pure state 100\=10\=1\=10 (because the complementary
orthogonal subspace is also orthogonal to 100\=10\=1\=10). Explicitly,
we have

\beq 2\times 100\10\1\10 = 1010\10\10 + 10\1010\10 - 00020000 -
00000200, \eeq
\beq \phantom{2\times 100\10\1\10}= 1100\1\100 + 1\1001\100 - 00020000 -
00000020, \eeq
\beq \phantom{2\times 100\10\1\10}= 11\1\10000 + 1\11\10000 - 00000200 -
00000020, \eeq
\beq \phantom{2\times 100\10\1\10}= 11\1\10000 + 00001\1\11 + 1010\10\10
- 01010101, \eeq
\beq \phantom{2\times 100\10\1\10}= 1\11\10000 + 00001\1\11 + 1100\1\100
- 00110011, \eeq
\beq \phantom{2\times 100\10\1\10}= 1100\1\100 + 1\1001\100 + 001\100\11
- 00110011, \eeq
\beq \phantom{2\times 100\10\1\10}= 1100\1\100 + 001\100\11 + 10\1010\10
- 01010101. \eeq

Quantum mechanics asserts that, for each one of the above equations,
there exists an experimental test which randomly selects one of the four
vectors appearing on the right hand side of that equation. A hidden
variable theory would claim that the selection is not random, and that
there are preassigned values, 0 and~1, corresponding to all 13 vectors:
on each line, one of the four vectors has value~1, the others have
value~0. Moreover, we may demand that, if the same vector appears in
different lines, the value associated to it is everywhere the same (that
is, the result of an experimental test which would determine that value
is not ``contextual''). These demands then lead us to the same
contradiction: there are 7 equations, so that the sum of values is 7. On
the other hand, each vector appears 2 or 4 times, so that the sum of
values is even. It is remarkable that no more than 13 propositions are
needed to reach the contradiction. The proof entirely holds in these
seven equations.

\bigskip Work by AP was supported by the Gerard Swope Fund, and the Fund
for Encouragement of Research.\bigskip

\frenchspacing
\begin{enumerate}
\item S. Kochen and E. P. Specker, J. Math. Mech. 17 (1967) 59.
\item M. Redhead, Incompleteness, nonlocality, and realism (Clarendon
Press, Oxford, 1987).
\item R. Clifton, Am. J. Phys. 61 (1993) 443.
\item H. Bechmann Johansen, Am. J. Phys. 62 (1994) 471.
\item A. Peres, J. Phys. A: Math. Gen. 24 (1994) L175.
\item A. Peres, Quantum theory: concepts and methods (Kluwer, Dordrecht,
1993) p. 114.
\item M. Kernaghan, J. Phys. A: Math. Gen. 27 (1994) L829.
\item N. D. Mermin, Phys. Rev. Lett. 65 (1990) 3373; Rev. Mod. Phys. 65
(1993) 803.
\item J. S. Bell, Physics 1 (1964) 195.
\item J. S. Bell, Rev. Mod. Phys. 38 (1966) 447.
\end{enumerate}

\newpage

\begin{center}{Table 1. \,Orthogonal octads used for proving the theorem.}

\bigskip\begin{tabular}{|c|ccccccccccc|}\hline\hline\str
 20000000 & & & & & & & & & & & \\ % removed
 02000000 & $\times$ & & & & & & & $\times$ & & & \\ %1
 00200000 & $\times$ & & & & & & $\times$ & & & & \\ %2
 00020000 & & & & & & & $\times$ & $\times$ & & & \\ %3
 00002000 & $\times$ & & & & & $\times$ & & & & & \\ %4
 00000200 & & & & & & $\times$ & & $\times$ & & & \\ %5
 00000020 & & & & & & $\times$ & $\times$ & & & & \\ %6
 00000002 & $\times$ & & & & & $\times$ & $\times$ & $\times$ & & & \\ %7
\hline\str
 11110000 & & $\times$ & & & & $\times$ & & & & & \\ %8
 11\=1\=10000 & & & & & & $\times$ & & & & $\times$ & \\ %9
 1\=11\=10000 & & & & & & $\times$ & & & $\times$ & & \\ %10
 1\=1\=110000 & & $\times$ & & & & $\times$ & & & $\times$ & $\times$ & \\ %11
 00001111 & & & & & & & & & & & \\ % removed
 000011\=1\=1 & & $\times$ & & & & & & & & $\times$ & \\ %12
 00001\=11\=1 & & $\times$ & & & & & & & $\times$ & & \\ %13
 00001\=1\=11 & & & & & & & & & $\times$ & $\times$ & \\ %14
\hline\str
 11001100 & & & & & & & $\times$ & & $\times$ & & \\ %15
 1100\=1\=100 & & & $\times$ & & & & $\times$ & & $\times$ & & $\times$ \\ %16
 1\=1001\=100 & & & $\times$ & & & & $\times$ & & & & \\ %17
 1\=100\=1100 & & & & & & & $\times$ & & & & $\times$ \\ %18
 00110011 & & & $\times$ & & & & & & $\times$ & & \\ %19
 001100\=1\=1 & & & & & & & & & $\times$ & & $\times$ \\ %20
 001\=1001\=1 & & & & & & & & & & & \\ % removed
 001\=100\=11 & & & $\times$ & & & & & & & & $\times$ \\ %21
\hline\str
 10101010 & & & & $\times$ & & & & $\times$ & & $\times$ & $\times$ \\ %22
 1010\=10\=10 & & & & & & & & $\times$ & & $\times$ & \\ %23
 10\=1010\=10 & & & & & & & & $\times$ & & & $\times$ \\ %24
 10\=10\=1010 & & & & & & & & & & & \\ % removed
 01010101 & & & & $\times$ & & & & $\times$ & & & \\ %25
 01010\=10\=1 & & & & & & & & & & $\times$ & $\times$ \\ %26
 010\=1010\=1 & & & & $\times$ & & & & & & $\times$ & \\ %27
 010\=10\=101 & & & & $\times$ & & & & & & & $\times$ \\ %28
\hline\str
 100101\=10 & $\times$ & & $\times$ & $\times$ & $\times$ & & & & & & \\ %29
 100\=10110 & $\times$ & $\times$ & $\times$ & & $\times$ & & & & & & \\ %30
 10010\=110 & $\times$ & & & & $\times$ & & & & & & \\ %31
 100\=10\=1\=10 & $\times$ & $\times$ & & $\times$ & $\times$ & & & & & & \\
%32
 0110\=1001 & & & & $\times$ & $\times$ & & & & & & \\ %33
 01\=101001 & & $\times$ & $\times$ & $\times$ & $\times$ & & & & & & \\ %34
 0\=1101001 & & $\times$ & & & $\times$ & & & & & & \\ %35
 0\=1\=10\=1001 & & & $\times$ & & $\times$ & & & & & & \\ %36
\hline\hline \end{tabular}\end{center}

\newpage

\begin{center}{\small Table 2. \,Construction of 14 planes from 20
vectors.}

\bigskip\begin{tabular}{|c|}\hline\hline\str
 02000000\qquad 00000002\qquad 00000020\\ % 1-7  7-6
\hline\str
 11110000\qquad 1\=1\=110000\qquad 00001\=1\=11\\ % 8-11  11-14
\hline\str
 1\=1001\=100\qquad 1100\=1\=100\qquad 001100\=1\=1\\ % 17-18  18-20
\hline\str
 01010101\qquad 10101010\qquad 01010\=10\=1\\ % 25-22  22-26
\hline\str
 10010\=110\qquad 100101\=10\qquad 01\=101001\qquad 0\=1101001\\
                                                  % 31-29  29-34  34-35
 0\=1\=10\=1001\qquad 100\=10110\qquad 100\=10\=1\=10\qquad 0110\=1001\\
                                                  % 36-30  30-32  32-33
\hline\hline \end{tabular}\end{center}

\end{document}